\documentclass[twocolumn,secnumarabic,amssymb,superscriptaddress,nobibnotes, aps, prd]{revtex4-2}

\usepackage{comment}
\usepackage{graphicx}
\usepackage[charter,greekuppercase=italicized]{mathdesign}
\usepackage{amsmath, amsfonts, amssymb}
\usepackage{hyperref}
\usepackage{xcolor}
\hypersetup{
    colorlinks = true,
    linkcolor = blue,
    anchorcolor = blue,
    citecolor = blue,
    filecolor = blue,
    urlcolor = blue
    }   
\usepackage{multirow}
\usepackage{hhline}
\DeclareUnicodeCharacter{2062}{\,}
\DeclareUnicodeCharacter{2009}{\,}
\DeclareUnicodeCharacter{03BC}{\,}
\DeclareUnicodeCharacter{1D445}{\,}

\begin{document}

\title{Quantum spin liquid ground state with the evidence of roton-like excitations at elevated temperatures in the triangular-lattice delafossite YbCuSe$_2$}

\author{K. Bhattacharya}
\address{Department of Physics, Shiv Nadar Institution of Eminence, Gautam Buddha Nagar, UP 201314, India}

\author{Y. Tokiwa}
\address{Advanced Science Research Center, Japan Atomic Energy Agency, Tokai, Ibaraki 319-1195, Japan}

\author{M. Majumder}
\email[Corresponding author: ]{mayukh.majumder@snu.edu.in}
\address{Department of Physics, Shiv Nadar Institution of Eminence, Gautam Buddha Nagar, UP 201314, India}

\date{\today}
\begin{abstract}
We present a comprehensive experimental investigation of the temperature evolution of magnetic states in triangular-lattice delafossite YbCuSe$_2$. Magnetization measurements on high-quality single crystals reveal easy-plane anisotropy. Specific heat, magnetization, and muon spin relaxation ($\mu$SR) establish the absence of magnetic order or spin freezing down to 0.03 K ($\leq J_{\mathrm{avg}}/250$), demonstrating a dynamically fluctuating quantum spin liquid (QSL) ground state. Thermodynamic measurements uncover multiple characteristic energy scales at $T_H \approx 4.5$ K, $T_L \approx 1.8$ K, and $T^* \approx 0.7$ K. Below $T^*$, $\mu$SR detects a dynamical phase separation in which the majority of the spins are forming a QSL state whereas the remaining spins form a sporadic, disorder-induced state decoupled from the dominant QSL component. Remarkably, the unconventional temperature dependence of the $\mu$SR relaxation rate indicates roton-like excitations emerging between $T_H$ and $T_L$, a feature not previously observed in any QSL system, preceding the stabilization of the low-temperature QSL at 0.3 K. These findings identify YbCuSe$_2$ as a unique QSL platform, providing valuable insights for further experimental and theoretical exploration. 
\end{abstract}

\maketitle

\textcolor{blue}{\textit{Introduction---}} Spin-1/2 triangular-lattice antiferromagnets have long attracted considerable interest as fertile platforms for exploring the interplay between geometric frustration and quantum fluctuations~\cite{doi:10.1139/p97-007}. A range of unconventional states have been proposed and realized in these systems, from nontrivial magnetic orders~\cite{Starykh_2015, PhysRevB.91.081104, PhysRevLett.114.027201, PhysRevLett.112.127203} to the highly sought-after quantum spin liquid (QSL) state. In QSL states, geometric frustration suppresses spontaneous symmetry breaking and stabilizes a dynamical ground state with fractionalized excitations~\cite{Balents2010, doi:10.1126/science.aay0668}. Numerous theoretical models have been developed to describe such QSL phenomena. Within the triangular-lattice Heisenberg $J_1$-$J_2$ model, a QSL state is predicted to emerge for $J_2/J_1$ ratios between 0.08 and 0.16~\cite{PhysRevB.93.144411, PhysRevB.107.165146, PhysRevLett.123.207203, PhysRevB.96.165141, PhysRevB.92.041105}. Including third nearest-neighbor exchange ($J_3$) in the $J_1-J_2$ Hamiltonian can even give rise to a chiral spin liquid (CSL) state~\cite{PhysRevB.92.140403,PhysRevB.100.241111}. For rare-earth-based systems, the $J_1$-$J_2$ XXZ model with easy-plane ($\Delta<1$) or easy-axis ($\Delta>1$) anisotropy is relevant, and a broad region in $J_2/J_1$ also supports QSL phases~\cite{PhysRevLett.134.196702}. Experimentally, the 4$f$-based delafossite family has emerged as a promising platform for observing QSL states~\cite{Bordelon2019, PhysRevB.101.224427, PhysRevB.100.224417, Scheie2024}, although the number of known compounds remains limited. Moreover, beyond geometric frustration, Kitaev-type exchange ($K$) frustration, involving anisotropic exchange couplings such as in-plane ($J_{\pm\pm}$) and off-diagonal ($J_{z\pm}$) terms~\cite{PhysRevResearch.7.023198, PhysRevLett.133.096703}, can stabilize QSL states~\cite{PhysRevLett.120.207203, PhysRevX.9.021017}. Experimentally, strong Kitaev exchange has been identified in long-range ordered delafossites such as CsCeSe$_2$ and KCeSe$_2$~\cite{PhysRevLett.133.096703, PhysRevResearch.7.023198}, but to date, a QSL state with dominant Kitaev interactions has not been observed.

Not only do exotic QSL ground states emerge at zero temperature, but multiple energy scales with distinct excitations have been predicted at finite temperatures~\cite{PhysRevB.99.140404, PhysRevB.110.214408, PhysRevResearch.2.013205}. Experimentally, these energy scales often manifest as a two-peak structure in magnetic heat capacity~\cite{PhysRevLett.62.1868, PhysRevLett.79.3451, PhysRevB.100.224417, Bordelon2019, doi:10.1126/science.1114727, PhysRevB.99.054421}, reported in several materials, including delafossite compounds~\cite{PhysRevB.100.224417, Bordelon2019}. Despite this, the microscopic nature of intermediate-temperature states remains unresolved. A prominent theoretical scenario is the appearance of roton-like gapped excitations (RLEs), analogous to vortex-like modes in superfluid He-II~\cite{PhysRev.94.262, PhysRev.113.1386, PhysRev.108.1346}. RLEs have also been experimentally observed in a few triangular-lattice systems, that ultimately develop magnetic order at low temperatures~\cite{Ito2017,Scheie2024, Li2020, PhysRevLett.133.186704}. Various mechanisms underpinning these excitations have been discussed~\cite{PhysRevB.73.174430, PhysRevLett.96.057201, DallaPiazza2015, PhysRevB.88.094407, PhysRevB.88.094407}. Interestingly, even in systems with QSL ground state is also proposed to have roton-like excitation with reduced gap compared to its magnetically ordered counterpart~\cite{PhysRevX.9.031026}, though remains unobserved experimentally. Thus, identifying roton-like excitations in systems with a QSL ground state is therefore intriguing, as it provides a unique platform to unravel how high-temperature excitation processes evolve toward a quantum-disordered regime at low temperatures. Such studies help clarify the relationship between finite-temperature dynamics and the stabilization of exotic quantum ground states in frustrated magnets.

\begin{figure}
\includegraphics[scale = 0.58]{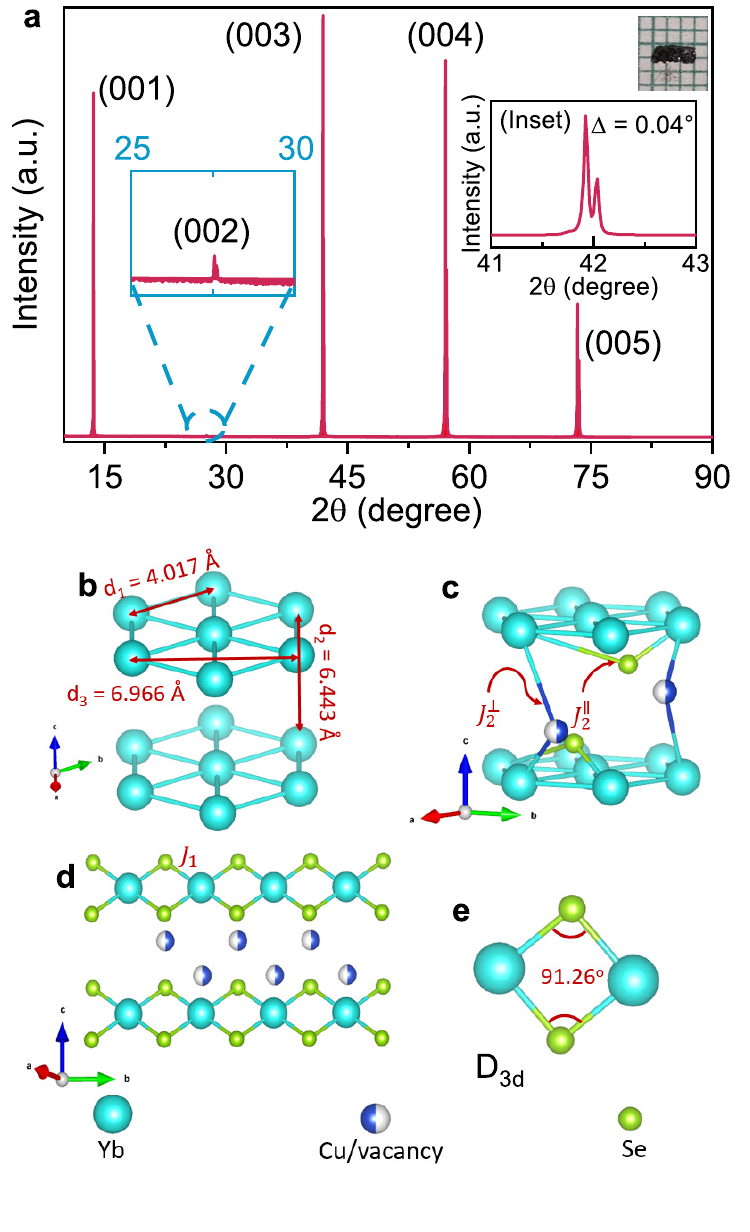}
\caption{\label{fig1} \textbf{a.} The XRD pattern for the (00l) plane displays sharp and well-defined peaks. The inset shows a representative peak. Splitting of the intensity lines corresponds to the $K_\alpha$ doublet, exhibiting the canonical intensity ratio of 2:1. \textbf{b.} Triangular lattice arrangement of Yb$^{3+}$ ions. \textbf{c.} Two possible superexchange pathways for the next-nearest-neighbor interaction are shown: $J_2^{\parallel}$ for in-plane, and $J_2^{\perp}$ for out of plane. The occupancy at the Cu-site is 0.5, indicating a random site disorder at the Cu site. \textbf{d.} The nearest-neighbor super exchange path ($J_1$). \textbf{e.} The Yb--Se--Yb superexchange pathway features a bond angle of 91.26$^\circ$.}
\end{figure}
 
This Letter reports the observations of multiple temperature scales in high-quality single crystalline YbCuSe$_2$, a novel 4$f$-based equilateral triangular-lattice system. Extensive thermodynamic and magnetic probes, including microscopic muon spin relaxation ($\mu$SR) experiments, reveal a QSL ground state. Most interestingly, unconventional temperature evolution of the $\mu$SR relaxation rate along with heat capacity provides evidence of RLEs before stabilizing the QSL state below 0.3~K, hitherto unobserved in a QSL system so far, placing YbCuSe$_2$ as a unique system.

\textcolor{blue}{\textit{Structural analysis---}} Millimeter-sized single crystals of YbCuSe$_2$ (a typical single crystal is shown in the inset of Fig.~\ref{fig1}a) were grown following the procedure described in Ref.~\cite{supplemental}. Single-crystal X-ray diffraction (XRD) confirms that YbCuSe$_2$ adopts the trigonal space group $P\overline{3}m1$ (No.~164)~\cite{supplemental}. High crystallinity is evidenced by the (00$l$) XRD pattern in Fig.~\ref{fig1}a, whose inset shows a full width at half maximum of intensity peak of just $0.04^\circ$, at par with the reported high-quality single crystals \cite{PhysRevB.108.024428,PhysRevLett.115.167203}. Structural analysis reveals that (Fig.\ref{fig1}b), Yb$^{3+}$ ions form an ideal triangular lattice with Yb--Yb nearest-neighbor distance, $d_{\mathrm{1}}=4.017(3)\,$\AA\,. In general, the next-nearest-neighbor interaction ($J_2$) represents a crucial superexchange pathway believed to play a key role in stabilizing a quantum spin liquid by suppressing conventional Néel order \cite{PhysRevB.93.144411}. For YbCuSe$_2$, the out-of-plane next-nearest-neighbor distance $d_2 = 6.443\,\text{\AA}$ (associated with the $J_2^{\mathrm{\perp}}$ interaction via a Yb-Cu-Yb superexchange path) is comparable to the in-plane next-nearest-neighbor distance $d_3 = 6.966\,\text{\AA}$ (associated with $J_2^{\mathrm{\parallel}}$ through a Yb-Se-Yb path), as illustrated in Fig.~\ref{fig1}c. In principle, the larger ionic radius of Se$^{2-}$ (184~pm) compared to Cu$^{1+}$ (91~pm)~\cite{https://doi.org/10.1107/S0567739476001551} enhances the polarizability of Se, allowing longer hopping pathways that can strengthen the in-plane superexchange $J_2^{\mathrm{\parallel}}$. However, since $d_2 < d_3$, it might possible that the out-of-plane coupling $J_2^{\mathrm{\perp}}$ is finite. Consequently, the energy scales of $J_2^{\mathrm{\parallel}}$ and $J_2^{\mathrm{\perp}}$ (corresponding to $d_3$ and $d_2$ respectively) are expected to be comparable, resulting in a delicate balance between these competing interactions \cite{Rojas_2011}. For YbCuSe$_2$, it is noteworthy that $J_2^\perp$ is mediated through the disordered Cu site (as shown in Fig.~\ref{fig1}c,d). Interestingly, for YbCuSe$_2$, the Yb--Se--Yb bond angle is $91.26^\circ$ as shown in Fig.~\ref{fig1}e, remarkably close to $90^\circ$ (closest among the reported 4$f$-based systems), which implies the possible presence of finite Kitaev exchange as have been seen in other delafossites even with a Yb--Se--Yb bond angle further away from $90^\circ$~\cite{PhysRevLett.133.096703, PhysRevResearch.7.023198}.

\begin{figure}[h]
\includegraphics[scale = 0.42]{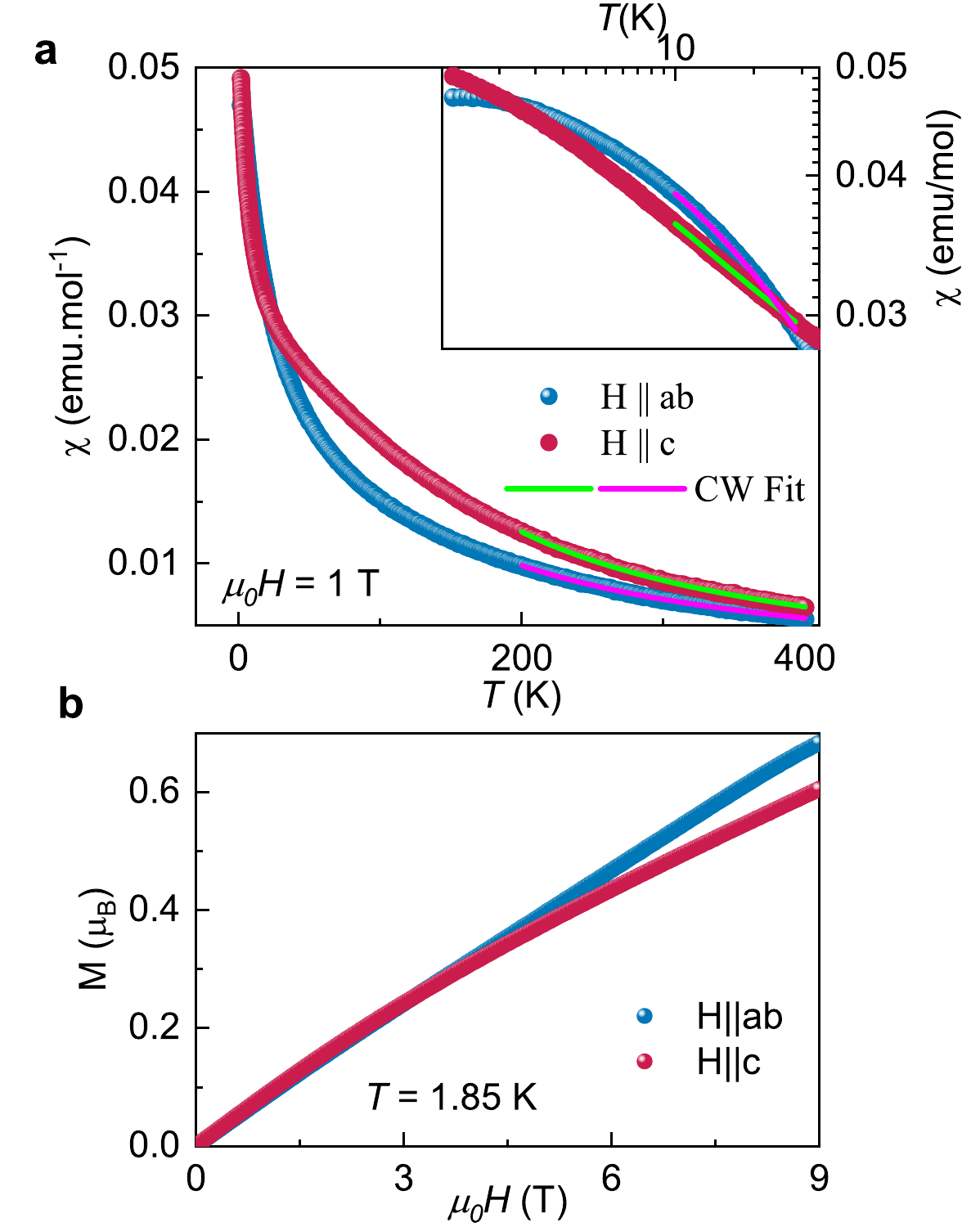}
\caption{\label{fig2} Magnetization of YbCuSe$_2$: \textbf{a.} \textit{dc} susceptibility ($\chi=M/H$) as a function of temperature measured at $\mu_0 H =1~$T for both directions, and the solid lines represent the CW fit. The inset depicts a low-temperature zoomed-in version of $\chi(T)$ along with the low-temperature CW fit. \textbf{b.} Magnetization as a function of magnetic field measured at $T=1.85~$K for both directions.}
\end{figure}

\textcolor{blue}{\textit{Magnetization---}} To obtain an estimation of the anisotropic exchange interactions strength in YbCuSe$_2$, d.c. magnetization measurements were carried out down to 1.8~K under applied fields parallel to the \textit{ab} plane and along the $c$ axis. The temperature dependence of the magnetic susceptibility $\chi(T)$ at $\mu_0 H=1$~T was analyzed using the Curie--Weiss (CW) expression $
\frac{1}{\chi(T) - \chi_0} = \frac{T - \theta_{\mathrm{CW}}}{C},$ where $C$ is the Curie constant, $\chi_0$ is the temperature-independent term, and $\theta_{\mathrm{CW}}$ is the CW temperature. Fits were performed in two regimes: a high-temperature range (200--400~K) and a low-temperature range (10--25~K), as shown in Fig.~\ref{fig2}a. The high temperature fits yields $\theta_{CW} = -48.51(6)~$K ($-35.75(2)~$K) and $\mu_{eff} =4.35(2) \mu_B$ ($4.98(3) \mu_B$) for the H$\parallel$c (H$\parallel$ab) direction. The $\mu_{eff}$ is close to the magnetic moment of the free Yb$^{3+}$-ion $ 4.53~\mu_B$. The low-temperature Curie--Weiss analysis (see inset of Fig.\ref{fig2}a) yields $\chi_{0,c}=0.0214(1)~$emu/mol, $\chi_{0,ab}=0.0053(4)~$emu/mol. $\theta_{c} = -8.47(2)$~K and $\theta_{ab} = -27.85(66)$~K, such a relatively high $\theta_{CW}$ has also been observed in other Delafossite compounds~\cite{10.1063/5.0071161, PhysRevMaterials.4.064410, PhysRevB.100.220407}. The negative sign of the $\theta_{CW}$ indicates an AFM-type of interaction in both directions. The $J$-values connected to the CW-temperatures, and can be calculated as $J_{zz}/k_B = 2 \theta_{c} /3 = - 5.65$~K  and $J_{\pm}/k_B = \theta_{ab} /3 = - 9.285$~K~\cite{PhysRevB.98.220409}, indicating the system exhibits an easy-plane anisotropy ($\Delta = J_{zz}/J_{\pm} = 0.61 <1$), and the average value could be $|J_{avg}/k_B|=\frac{2 |J_{\pm}| + |J_{zz}|}{3k_B } = 8.07~$K. The effective magnetic moments of $\mu_{\mathrm{eff}, {c}} = 1.45(8) \,\mu_B$ and $\mu_{\mathrm{eff}, {ab}} = 3.17(2) \,\mu_B$, both reduced relative to the free-ion value of Yb$^{3+}$, consistent with strong spin-orbit coupling and CEF renormalization. 

The magnetization isotherm measured at 1.85~K as a function of applied magnetic field, as depicted in Fig.~\ref{fig2}b, which shows even at the highest measured field of 9~T, the magnetic moment reaches only approximately 0.8\,$\mu_B$ for both directions without saturation. This suggests a higher field is required to achieve saturation, similar observations found in other delafossite compounds~\cite{PhysRevB.99.180401, 10.1063/5.0071161}.

\begin{figure}
\includegraphics[scale = 0.36]{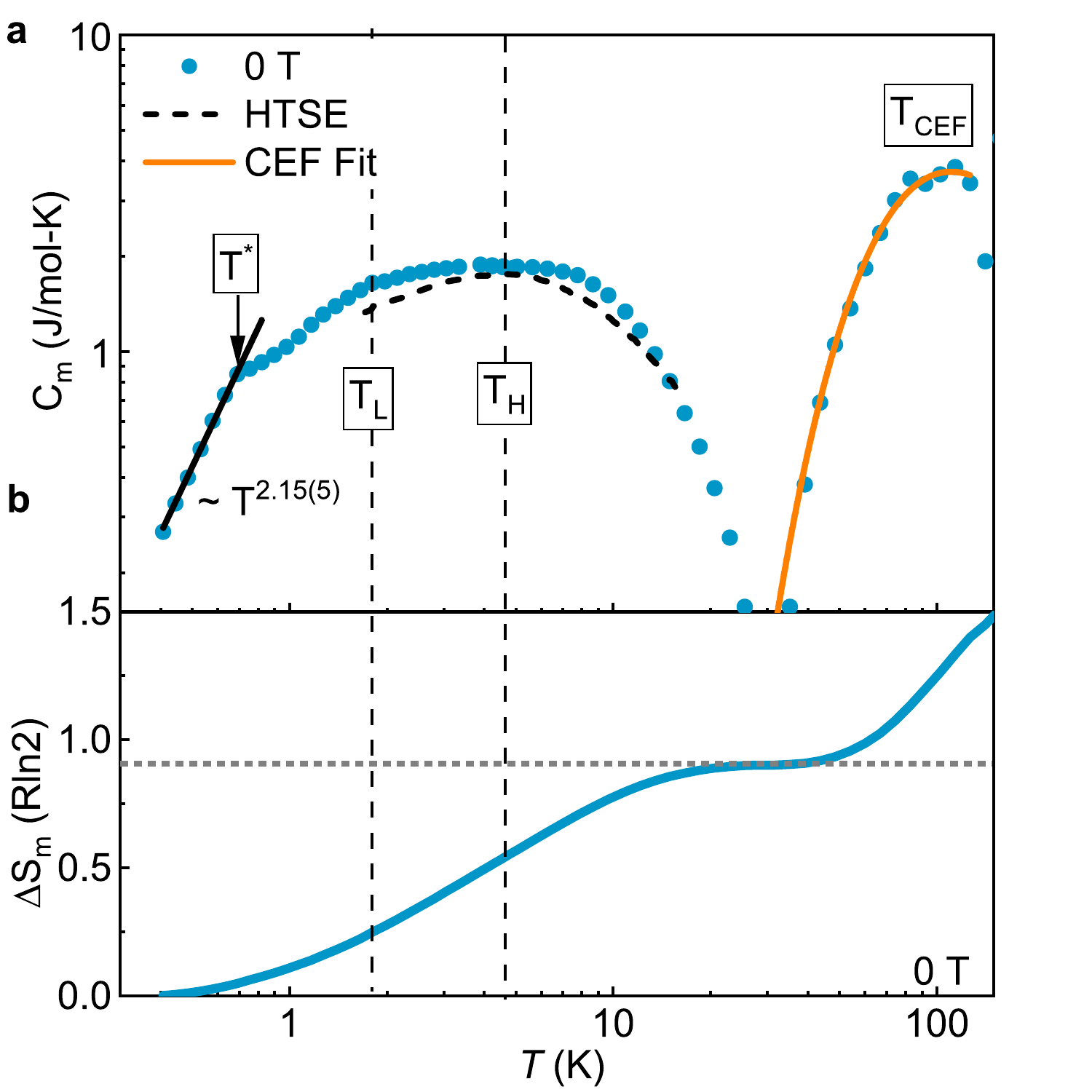}
\caption{\label{fig3} 
\textbf{a.} Magnetic specific heat ($C_m$) of YbCuSe$_2$ as a function of temperature measured at $\mu_0 H = 0$~T. The black solid line shows a power-law dependence of $\sim T^{2.15(5)}$ below $T^* = 0.7$~K. The black dashed lines represent the high-temperature series expansion (HTSE) for a Heisenberg S=1/2 triangular lattice antiferromagnet, adopted from Ref.~\cite{PhysRevLett.71.1629} with $J = 8.18\,\mathrm{K}$. The orange solid line represents the Schottky fit corresponding to the Crystal electric field (details are in the Ref.\cite{supplemental}). \textbf{b.} The change of magnetic entropy as a function of temperature at $\mu_0 H = 0$~T. 
}
\end{figure}

\textcolor{blue}{\textit{Specific heat---}} To elucidate the low-energy excitations in YbCuSe$_2$, heat capacity measurements were performed down to 0.4~K. Subtracting the phonon background using non-magnetic analog LuCuSe$_2$, the magnetic heat capacity $C_m(T)$ of YbCuSe$_2$ has been obtained, depicted in Fig.~\ref{fig3}a. Consistent with susceptibility results, the absence of sharp anomalies in $C_m(T)$ rules out long-range magnetic order (down to 0.4~K), instead, shows the presence of several energy scales with lowering temperature. $C_m(T)$ show a high-temperature broad peak at $T_H \approx 4.5$~K and two lower-temperature features at $T_L \approx 1.8$~K and $T^* \approx 0.7$~K. Theoretically, the characteristic temperatures $T_H$ and $T_L$ are intimately linked to the underlying exchange interactions, with a relation $T_L/J \sim 0.2$ and $T_H/J \geq 0.5$ \cite{PhysRevB.101.235115,PhysRevB.99.140404, PhysRevB.98.035107, PhysRevX.8.031082}, and also observed experimentally in other triangular lattices \cite{PhysRevB.100.224417, Bordelon2019}. Using the average exchange value $|J_{\mathrm{avg}}|/k_B = 8.07$~K, as determined from CW analysis, the theoretically expected values are $T_L = 1.61$~K and $T_H = 4.44$~K, both in excellent agreement with the observed temperatures in the heat capacity data. 

\begin{figure*}
\includegraphics[scale = 0.35]{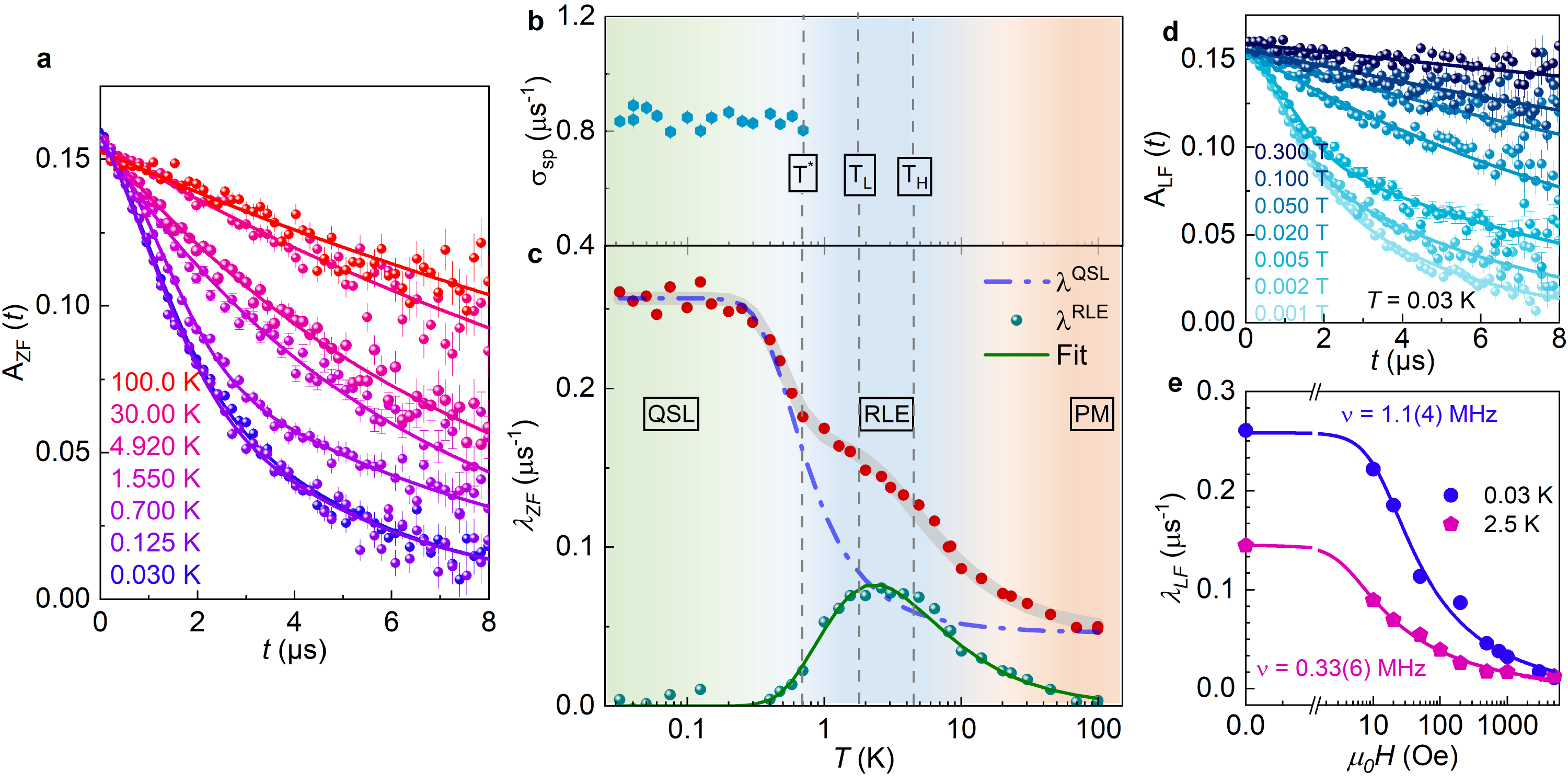}
\caption{\label{fig4} 
\textbf{a.} Muon asymmetry as a function of time at various temperatures under zero-field conditions; solid lines represent theoretical fits as described in the text.
\textbf{b.} $\sigma_{sp}$ as a function of temperature appears at $T<T^*$.
\textbf{c.} Zero-field relaxation rate $\lambda_{\mathrm{ZF}}$ as a function of temperature; the blue dashed-dot line is the fit corresponding to the QSL states mentioned in the main text. The green circles represent the contribution of the RLE. Green solid line corresponds to an empirical function $\lambda^{\mathrm{RLE}}(T) = \left( \frac{a}{T} \right) \exp \left( -\frac{\delta}{T} \right),$ where $a$, and $\delta$ are the constant and energy gap, respectively. The grey solid line is for the guide to the eyes. 
\textbf{d.} Muon asymmetry versus time for different applied longitudinal fields measured at $T = 0.03$~K. Solid lines are theoretical fits mentioned in the Ref.\cite{supplemental}.
\textbf{e.} Longitudinal relaxation rate $\lambda_{\mathrm{LF}}$ as a function of applied magnetic field for $T = 0.03$~K and $T = 2.5$~K; solid lines represent the theoretical fits discussed in the Ref.\cite{supplemental}.}
\end{figure*}

Furthermore, the maximum of the magnetic heat capacity ($C_m^{\mathrm{max}}$) at $T_H$ is found to be $C_m^{\mathrm{max}} \approx 0.22 R$, which closely matches the theoretical prediction for frustrated isotropic triangular antiferromagnets~\cite{PhysRevB.63.134409}. The high-temperature series expansion (HTSE) also reproduces the temperature dependence reasonably (as shown in Fig.~\ref{fig3}a), with an exchange coupling $J = 8.18\,\mathrm{K}$, consistent with the average exchange $J_{\mathrm{avg}}$ of YbCuSe$_2$~\cite{PhysRevLett.71.1629}. However, achieving a better fit may require incorporating higher-order interactions and anisotropic terms. The magnetic entropy change 
$\Delta S_m(T) = \int_{0.4~\mathrm{K}}^{T} \frac{C_m(T')}{T'}\,dT'$  
approaches 90\% of the expected $R\ln2$ at around 25~K (Fig.~\ref{fig3}b), substantiating a well-isolated $J_{\mathrm{eff}}=1/2$ Kramer's doublet ground state. Theoretically, the thermal entropy per site at the low-temperature peak (here denoted as $T_L$) is expected to reach approximately 1/3 of the high-temperature limit $R\ln2$ \cite{PhysRevB.99.140404, PhysRevLett.134.226701}. This prediction is well supported by the experimental data of YbCuSe$_2$, as the entropy released at $T_L$ is about $0.33 \times R\ln2$ (by considering that the remaining 10\% entropy will be released below $T<0.4~$K), in excellent agreement with theoretical expectations. After establishing the consistency of the experimentally observed energy scales $T_H$ and $T_L$ with theoretical predictions, it is important to emphasize that roton modes are expected to be thermally activated in the temperature window between $T_H$ and $T_L$~\cite{PhysRevB.99.140404, PhysRevB.110.214408}. 

A quadratic temperature dependence ($C_m \propto T^{2.15}\approx T^2$) below $T^* = 0.7$~K is exhibited (Fig.~\ref{fig3}a). Notably, such a power-law exponent has also been predicted for a gapless Dirac- or Nodal-like QSL state~\cite{PhysRevLett.98.117205} and also experimentally observed in Dirac QSL candidates~\cite{PhysRevLett.133.266703, Bordelon2019, Wu2022, PhysRevLett.125.267202}. An onset of the disorder-induced phase, which consists of 27\% of the spins, coincides with this weak kink at $T^*$ (as evidenced from $\mu$SR, discussed in the next section). Thus, in the temperature dependence of $C_m$ below $T^*$, the disorder-induced phase has a subdominant contribution, heat capacity being a volume-sensitive probe. However, disorder-induced state (e.g., random valence bond) is expected to show a sub-linear power-law~\cite{PhysRevX.8.031028, Kimchi2018}, in contrast to the present findings for YbCuSe$_2$, further signaling that $C_m(T)$ below $T^*$ is not solely dominated by the disorder.

\textcolor{blue}{\textit{Muon spin relaxation ($\mu$SR)---}}As described above, multiple characteristic temperature scales have been identified from the magnetic heat capacity, 
$C_m(T)$, indicating possible crossovers between distinct dynamical states and roton-like excitations between $T_H$ and $T_L$. To get further microscopic nature of the temperature evolution of the states, we have employed muon spin relaxation ($\mu$SR)-- a highly sensitive microscopic probe, which effectively distinguishes static or dynamic correlations. The zero-field (ZF) asymmetry spectra are displayed in Fig.~\ref{fig4}a. At the base temperature of 0.03~K, the spectra exhibit neither an oscillatory nor initial asymmetry drop-- a characteristic of long-range magnetic order, nor a 1/3 recovery tail typically associated with spin-freezing. In the high-temperature paramagnetic (PM) regime, the spectra are well described by a single exponential relaxation function, $
A_{\mathrm{ZF}}(T > T^*) = A_0 e^{-\lambda_{\mathrm{ZF}} t},
$
Where $A_0$ and $\lambda_{\mathrm{ZF}}$ denote the initial asymmetry and the ZF relaxation rate, respectively. At lower temperatures ($T < T^*$), a single exponential relaxation no longer captures the spectra; rather, a two-component fractional weighting function is required:
$
A_{\mathrm{ZF}}(T < T^*) = A_0 \left[ \tilde{f} e^{-\lambda_{\mathrm{ZF}} t} + (1-\tilde{f}) e^{-\frac{1}{2}(\sigma_{\mathrm{sp}} t)^2} \right],
$
where $\sigma_{\mathrm{sp}}$ and $\tilde{f}$ represent the Gaussian relaxation rate and fractional contribution, respectively. The Gaussian term accounts for a pronounced early-time hump-like feature in the spectra. A temperature-independent value of $\sigma_{\mathrm{sp}} \approx 0.85~\mu\mathrm{s}^{-1}$ with a temperature independent fraction $(1-\tilde{f}) = 0.27$ is obtained, indicating a persistent broadening below 0.7~K (see Fig.\ref{fig4}b). Such a behavior is associated with a sporadic state, as has been seen in other disordered frustrated systems~\cite{PhysRevLett.73.3306,PhysRevLett.127.157204}. It should be mentioned that appearance of this disorder-induced state is not unexpected for a compound with structural disorder at the Cu site (as shown in Fig.\ref{fig1}). Also note that the muon site calculation indicates a single muon site close to the vacancy site~\cite{supplemental}, indicating the appearance of magnetic phase separation below 0.7~K. Thus, approximately 27\% of the spins participate in forming the disorder-induced state, whereas, interestingly, the remaining 73\% experience homogeneous fluctuating internal fields ($\beta = 1$)\cite{streched_exponent}, as reflected in the exponential component $\lambda_{\mathrm{ZF}}$ down to 0.03~K. It is also interesting to point out that similar magnetic phase separation has been reported in several structurally ordered delafossite compounds, where there is a coexistence of dynamically fluctuating phase with short-ranged magnetically ordered or spin-glass state~\cite{Scheie2024, Xie2023, TlYbSe2, PhysRevB.103.144413}.

Taking advantage of $\mu$SR being a microscopic tool, we were able to disentangle the disorder-free contribution and the disorder-induced contributions (as discussed above). Let us now discuss the temperature evolution of the $\lambda_{\mathrm{ZF}}$ (associated with the disorder-free phase even below $T^*$). At high temperatures ($\gtrsim$ 30 K), $\lambda_{\mathrm{ZF}}$ remains essentially constant, characteristic of a paramagnetic regime with fast spin fluctuations~\cite{supplemental}. Upon cooling, $\lambda_{\mathrm{ZF}}$ starts to increase due to the slowing down of the spin fluctuations, consistent with the enhancement of $C_m$ from a similar temperature range (see Fig.\ref{fig4}c). With further decrease in temperature, $\lambda_{\mathrm{ZF}}$ develops a "knee"-like structure for $T_L<T<T_H$. Below T$_L$, $\lambda_{ZF}$ increases again and saturates below 0.3~K, signaling persistent spin dynamics expected for a QSL state~\cite{PhysRevLett.73.3306, PhysRevLett.109.037208}. Notably, the overall temperature dependence of $\lambda_{\mathrm{ZF}}$, from the PM regime to the QSL state, reveals an unconventional evolution with an intermediate phase, contrasting sharply with the typical behavior reported for other QSL systems~\cite{PhysRevB.110.L060403, Arh2022, PhysRevB.105.094439} whose temperature dependence can be empirically modeled by $\lambda^{QSL} (T) = \frac{\lambda_{0}}{1+ \eta ~ exp(- T_{s} / T)}$ (depicted by the blue dashed-dot line in Fig.~\ref{fig4}c and also its deviation from the experimental data points supports the unusual temperature dependence). The parameter $\lambda_0$ is the constant value at which $\lambda_{\mathrm{ZF}}$ saturates, while $T_S$ is a characteristic energy scale and $\eta$ is the exponential prefactor~\cite{supplemental}. 

To further probe the dynamical character of these phases, we carried out longitudinal field $\mu$SR measurements at two specific temperatures: 2.5 K (within the intermediate regime between $T_H$ and $T_L$) and 0.03 K, corresponding to the QSL state (Fig.~\ref{fig4}d). Even when exposed to a strong longitudinal field of 0.3~T, the muon relaxation is not quenched, demonstrating highly dynamic correlations. The extracted fluctuation rates are $\nu = 1.1~\mathrm{MHz}$ and $\nu = 0.3~\mathrm{MHz}$, for 0.03~K and 2.5~K respectively (see Fig.\ref{fig4}e and details are given in the Ref.\cite{supplemental}), comparable to those found in other QSL systems~\cite{PhysRevB.100.241116, PhysRevLett.125.267202}.

Having established the presence of an intermediate phase between $T_L$ and $T_H$ above the QSL ground state, we now turn to the mechanisms governing this regime. Considering spin-1/2 triangular-lattice, theory predicts the emergence of gapped roton-like excitations (RLEs) at elevated temperatures, where a local minimum in the dynamical structure factor develops at a specific wave vector with a gap in the range between $0.5J$ to $0.8J$ for magnetically ordered ground states, depending on the specific microscopic model~\cite{PhysRevB.79.144416, PhysRevB.99.140404, PhysRevB.102.064421, PhysRevB.88.094407, PhysRevB.74.224420}. Also, in the temperature range between $T_L$ and $T_H$, the contribution of the wave vector related to the RLE dominates~\cite{PhysRevB.99.140404, PhysRevResearch.2.013205}. To extract out the RLE contribution, we evaluate $\lambda^{\mathrm{RLE}}(T)=\lambda_{\mathrm{ZF}}(T)-\lambda^{\mathrm{QSL}}(T),$ as shown in Fig.~\ref{fig4}c. The resulting hump in $\lambda^{\mathrm{RLE}}(T)$ appears between $T_L$ and $T_H$. The RLE gap $\delta$, extracted using 
$\lambda^{\mathrm{RLE}}(T) = \left( \frac{a}{T} \right)\exp\!\left(-\frac{\delta}{T}\right)$ (solid line in Fig.~\ref{fig4}c), is found to be $2.3$~K ($\simeq 0.3J_{\mathrm{avg}}$). Interestingly, the reduced gap scale is also supported by theoretical study where a softening of the RLE gap is expected in systems hosting a QSL ground state compared to their magnetically ordered counterparts~\cite{PhysRevX.9.031026}. Hence, in YbCuSe$_2$, RLEs dominate the intermediate-temperature regime, while at lower temperatures QSL-related excitations become prominent and saturate below 0.3\,K. RLEs may originate from several microscopic mechanisms---including vortex--antivortex fluctuations~\cite{PhysRevB.73.174430}, spinon--antispinon pairing~\cite{PhysRevLett.96.057201, DallaPiazza2015}, or interaction-stabilized magnon modes~\cite{PhysRevB.88.094407, PhysRevB.110.214408}. Among these, spinon--antispinon pairing appears most relevant for a QSL-candidate system such as YbCuSe$_2$. Nevertheless, resolving the precise nature of the RLEs will require complementary probes such as neutron scattering and NMR.

Furthermore, it is to be mentioned that, the saturation of $\lambda_{\mathrm{ZF}}$ starts below 0.3~K and is featureless at $T^*$, whereas the presence of the sporadic phase (related to disorder) starts below $T^*$, suggesting that the sporadic spins (about 27\%) are not coupled to the main QSL phase unlike the systems with magnetic-site disorder~\cite{PhysRevLett.127.157204, PhysRevB.83.180416}, which is further supported by the fact that $\beta$ remains unity (across $T^*$) down to the lowest temperature ($\sim$0.03~K). 

\textcolor{blue}{\textit{Conclusion-}} We have presented a comprehensive investigation of the temperature evolution of the new triangular-lattice delafossite YbCuSe$_2$. Magnetization measurements on high-quality single crystals yield $J_{zz}/k_B = -5.65$ K and $J_{\pm}/k_B = -9.29$ K, establishing easy-plane anisotropy with $\Delta = J_{zz}/J_{\pm} = 0.61 < 1$. Specific heat, magnetization, and microscopic $\mu$SR measurements collectively demonstrate the absence of magnetic order down to 30 mK ($\leq J_{\rm avg}/250$), confirming a dynamical QSL ground state. The magnetic heat capacity $C_m(T)$ reveals multiple characteristic energy scales, $T_H \approx 4.5$ K, $T_L \approx 1.8$ K, and $T^* \approx 0.7$ K, upon cooling. $\mu$SR further uncovers a dynamical phase separation below $T^*$: a minority of the spins form a sporadic state rooted in non-magnetic site disorder, while the majority of the spins form the QSL state. Importantly, these two components remain effectively decoupled, in stark contrast to behavior typically observed in systems with magnetic-site disorder. 

Most notably, we have observed the energy scales $T_H$ and $T_L$ in heat capacity and an unconventional temperature dependence of the $\mu$SR relaxation rate in the intermediate regime $T_L < T < T_H$ before the system stabilizes into the QSL state below 0.3 K. The magnitudes of $T_H$ and $T_L$, along with the gap estimated from the $\mu$SR relaxation rate, are consistent with theoretical predictions for gapped RLEs in a spin-1/2 triangular lattice. The observation of RLEs in a system with QSL ground state is unique in YbCuSe$_2$. Our results, therefore, motivate further theoretical studies and utilization of complementary probes, to unravel the full landscape of emergent states in YbCuSe$_2$. YbCuSe$_2$ thus establishes itself as a benchmark triangular-lattice QSL candidate, distinguished by the emergence of roton-like excitations.

\textcolor{blue}{\textit{Acknowledgment---}} We acknowledge H. Luetkens, PSI, Switzerland, and I. Ishant, SNIoE, India, for their help during the $\mu$SR measurements.

\textcolor{blue}{\textit{Note added.}} During the manuscript preparation, we became aware of Refs.~\cite{m33y-7fjb, 5h6v-nm9s} where bulk measurements have been reported on the same compound.


\bibliography{references}
\end{document}